\documentclass[11pt,english]{article}
\usepackage[T1]{fontenc}
\usepackage[latin9]{inputenc}
\usepackage{datetime}
\usepackage{color}
\usepackage{amsthm}
\usepackage{amsmath}
\usepackage{amssymb}
\usepackage{cite}

  \theoremstyle{plain}
  
  \theoremstyle{plain}
  \newtheorem*{lem*}{Lemma}
\theoremstyle{plain}
\newtheorem{thm}{Theorem}

\newcommand{\be}{\begin{equation}}
\newcommand{\ee}{\end{equation}}
\newcommand{\kb}[2]{|#1\rangle\langle#2|}
\newcommand{\bk}[2]{\langle#1|#2\rangle}
\newcommand{\ket}[1]{|#1\rangle}
\newcommand{\bra}[1]{\langle#1|}
\newcommand{\C}{\mathbb{C}}
\newtheorem{lemma}{Lemma}

\renewcommand{\baselinestretch}{1.1}
\usepackage{amsfonts}
\usepackage{amsthm}
\usepackage[labelfont=bf, font=small]{caption}

\usepackage{babel}

\begin{document}

\title{\large\bf{On the Impossibility to Extend Triples of\\
Mutually Unbiased Product Bases in Dimension Six}}
\author{Daniel McNulty and Stefan Weigert\\
 Department of Mathematics, University of York\\
York YO10 5DD, UK\\ \\
\small{\tt{dm575@york.ac.uk, stefan.weigert@york.ac.uk}}}
\date{1 October 2012}
\maketitle


\begin{abstract}
An analytic proof is given which shows that it is impossible to extend any triple of mutually unbiased (MU) product bases in dimension six by a single MU vector. Furthermore, the 16 states obtained by removing two orthogonal states from any MU product triple cannot figure in a (hypothetical) complete set of seven MU bases. These results follow from exploiting the structure of MU product bases in a novel fashion, and they are among the strongest ones obtained for MU bases in dimension six \emph{without} recourse to computer algebra.  
\end{abstract}

\section{Introduction}

Two orthonormal bases $\mathcal{B}=\{\ket{\phi_1},\ldots,\ket{\phi_d}\}$ and $\mathcal{B}'=\{\ket{\psi_1},\ldots,\ket{\psi_d}\}$ of the $d$-dimensional Hilbert space $\mathbb{C}^d$ are \emph{mutually unbiased} (MU) if and only if $|\bk{\psi_j}{\phi_k}|^2=1/d$ for all $j,k=1\ldots d$ \cite{schwinger+60}. This property expresses the notion of \emph{complementarity} for discrete variables. Not surprisingly, MU bases have a number of useful applications. For example, they provide an optimal means to reconstruct unknown quantum states \cite{wootters+89}, and they allow one to establish secret keys useful for cryptographic ends \cite{cerf+02}.

Much is known about MU bases of quantum systems if the number $d$ of discrete levels equals a prime or prime power, i.e. $d=p^n$ (cf. \cite{durt+10}). In these cases, \emph{complete sets} of $(p^n+1)$ MU bases can be constructed \cite{wootters+89,ivanovic+81} while their (non-) existence for ``composite'' dimensions such as $d=6,10,12\ldots$ remains unknown.

Considerable efforts have been devoted to dimension six, the smallest dimension where a complete set is elusive. Triples of MU bases have been constructed in a number of ways \cite{bengtsson+07,Jaming+2009,brierley+09,aschbacher+07} but only piecemeal progress has been made towards a proof of the conjecture that only three MU bases exist for $d=6$ \cite{zauner+99}. For example, increasingly strong numerical evidence \cite{butterley+07,brierley+08} supports this view, and the use of computer-algebraic methods \cite{grassl+04,brierley+09,raynal+11} allows one to rule out certain MU bases from being part of a complete set. 

Only a few \emph{analytic} results are known for sets of MU bases in composite dimensions such as $d=6$ or $d=10$. Let us briefly summarize them: (i) complete sets of MU bases are equivalent to orthogonal decompositions of Lie algebras \cite{boykin+07,kibler08}; (ii) in specific composite dimensions, \emph{more} than $(q^a+1)$ MU bases can be constructed using Latin squares \cite{wocjan+05} -- here $q^{a}$ is the smallest factor in the prime decomposition of $d$; (iii) given a ``nice unitary error basis'' in dimension six, none of its partitions gives rise to more than three MU bases \cite{aschbacher+07}; (iv) a complete set of MU bases contains a fixed amount of entanglement which implies that, in dimension six, any such set will contain at most three MU \emph{product} bases \cite{wiesniak+11}; (v) an exhaustive list of inequivalent \emph{pairs} and \emph{triples} of MU \emph{product} bases has been established in dimension six \cite{mcnulty+11}; (vi) no complete set of MU bases in dimension six contains both the standard basis and the Fourier basis \cite{matolcsi+12}.

The derivation of the analytic result presented in this paper hinges on the \emph{product} structure of the Hilbert space $\mathbb{C}^6$. In some sense, our approach extends successful studies of entanglement in prime power dimensions where the product structure of MU bases plays an important role \cite{romero+05,Lawrence11}. However,  we will obtain a \emph{negative} result: \emph{in dimension six, no triple of mutually unbiased product bases can be extended by a single MU vector}. This result is already known -- however, the final step of the  recent proof in \cite{mcnulty+11} relies on algebraic manipulations carried out by a computer \cite{grassl+04}. In contrast, the method presented here is entirely analytic and elementary.

This paper is set out as follows. We start Sec. 2 by recalling the list of all MU product triples in dimension six, followed immediately by a proof of the main theorem. In Sec. 3 we marginally improve the theorem by introducing MU product constellations and discuss the limitations of this approach. We summarise and discuss our results in Sec. 4.

\section{MU product triples in $d=6$}

\subsection{All MU product triples}

The starting point of our derivation is the fact that, in dimension six, no more than two triples of MU \emph{product} bases exist which are inequivalent under specific unitary or anti-unitary transformations, defined by the requirements to respect the product structure of the states and to leave invariant the modulus of their inner products, as explained in \cite{mcnulty+11}. The  transformations include a unitary map acting on all bases simultaneously, the multiplication of any state by an arbitrary phase factor, the permutation of states within a basis, and the complex conjugation of all bases; in addition one can re-order the bases arbitrarily.

The following lemma lists all triples of MU product bases, up to equivalence defined by the transformations just described.    

\begin{lemma}\label{lemma1}
\label{thm:two-triples-d=6} Any triple of MU product bases
in the space $\mathbb{C}^{2}\otimes\mathbb{C}^{3}$ is equivalent
to either  
\begin{align}
\mathcal{T}_{0} &
=\{\ket{j_{z},J_{z}};\,\ket{j_{x},J_{x}};\,\ket{j_{y},J_{y}}\}\,,
 \nonumber \\
\mbox{or }\quad \mathcal{T}_{1} &
=\{\ket{j_{z},J_{z}};\,\ket{j_{x},J_{x}};\,\ket{0_{y},J_{y}},\ket{1_{y},J_{w}}\}\,,\end{align}
where $j=0,1$ and $J=0,1,2$.
\end{lemma}

The bases in both triples are expressed in terms of states which form complete sets of MU bases in the spaces $\C^2$ and $\C^3$. We denote the complete set of three MU bases in $\C^2$ by $\{\ket{j_z}\}$, $\{\ket{j_x}\}$ and $\{\ket{j_y}\}$, where $j=0,1$, and the complete set of four MU bases in $\C^3$ is given by ${\cal B}_k \equiv \{\ket{J_k}, J=0,1,2\}, k=x,y,z,w$. These bases consist of the eigenstates of the Heisenberg-Weyl operators 
$Z$, $X$, and $Y\equiv XZ$ (for $\C^2$), and of $Z$, $X$, $Y$, and $W\equiv X^2Z$ (for $\C^3$) \cite{bandyo+01}. The operators $X$ and $Z$ satisfy $ZX=\omega XZ$, with $\omega=e^{2\pi i/d}$ for $d=2,3$, i.e. they are the cyclic shift (modulo $d$) and phase operators, respectively. For the sake of brevity, we denote the six orthogonal product states $\ket{j_z}\otimes\ket{J_z}$, with $j=0,1$, and $J=0,1,2$, by $\ket{j_z,J_z}$.
 
Notice that the third basis of $\mathcal{T}_1$ contains states from \emph{two} bases of $\C^3$, namely ${\cal B}_y\equiv\{\ket{J_y}\}$ and ${\cal B}_w\equiv\{\ket{J_w}\}$ so that it \emph{cannot} be written as a direct product of a basis in $\C^2$ with a basis in $\C^3$; following \cite{wiesniak+11}, we call such a basis \emph{indirect}. The third basis of $\mathcal{T}_0$ is a \emph{direct} product basis since it \emph{can} be written in that particular form.

For later reference we now write out the matrix representations of the MU bases ${\cal B}_y$ and ${\cal B}_w$ in the computational basis,
\begin{equation}
H_{y}=\frac{1}{\sqrt{3}}\left(\begin{array}{ccc}
1 & 1 & 1\\
\omega & \omega^{2} & 1\\
\omega & 1 & \omega^{2}\end{array}\right)\quad\mbox{and}\quad
H_{w}=\frac{1}{\sqrt{3}}\left(\begin{array}{ccc}
1 & 1 & 1\\
\omega^{2} & 1 & \omega\\
\omega^{2} & \omega & 1\end{array}\right)\,.\label{yandwmatrix}\end{equation}
The columns of each matrix are orthogonal, and since the modulus of each entry equals $1/\sqrt{3}$, the matrices $H_y$ and $H_w$ are complex $(3\times3)$ Hadamard matrices. Thus, the complete set of MU bases in dimension $d=3$ can be written as the set of $(3\times3)$ matrices $\{I,F_3,H_y,H_w\}$, where $I$ is the identity and $F_3$ is the Fourier matrix,
\begin{equation}
F_3\equiv\frac{1}{\sqrt{3}}\left(\begin{array}{ccc}
1 & 1 & 1\\
1 & \omega & \omega^{2}\\
1 & \omega^{2} & \omega\end{array}\right).
\end{equation}

\subsection{Excluding MU product triples from a complete set of MU bases}

We now present the main result of this paper, which is an analytic proof of the following theorem.

\begin{thm}
\label{thm:unextendible-triple}
No triple of MU product bases in dimension six can be extended by a single MU vector.
\end{thm}

In other words, any triple of MU product bases acts like a \emph{cul-de-sac} when attempting to construct a complete set of MU bases in dimension six. The first proof of this result in \cite{mcnulty+11} depends on an exact computer-algebraic search: Lemma \ref{lemma1} guarantees that any triple of MU product bases contains the pair $\{\ket{j_{z},J_{z}}\}$ and $\{\ket{j_{x},J_{x}}\}$ (or is equivalent to such a pair) and, according to \cite{grassl+04}, only 48 vectors exist which are MU to this pair . However, there is no subset of these 48 vectors which, when combined with $\{\ket{j_{z},J_{z}}\}$ and $\{\ket{j_{x},J_{x}}\}$, would give rise to more than a \emph{triple} of MU bases. Consequently, no state can be MU to either $\mathcal{T}_{0}$ or $\mathcal{T}_{1}$ which implies Theorem 1.

We now proceed to prove Theorem \ref{thm:unextendible-triple} \emph{analytically}. Due to Lemma \ref{lemma1}, it is sufficient to show that no vector is MU to either of the triples $\mathcal{T}_{0}$ or $\mathcal{T}_{1}$; any other MU product triple can be transformed to one of these two triples using the equivalence transformations described above. 

A candidate state $\ket{\psi} \in \mathbb{C}^6$ is MU to the three product bases $\mathcal{T}_0$ if and only if the following 18 conditions hold, 
\begin{equation}
\left\vert \bk{j_a,J_a}{\psi}\right\vert ^{2}=\frac{1}{6}\, ,\quad a=x,y,z\, ;j=0,1\, ;
J=0,1,2\, ,
\label{eq: c6 conditions on psi}
\end{equation}
not all of which are independent. Similarly, the state $\ket{\psi}$ is mutually unbiased to the product triple $\mathcal{T}_1$ if and only if
\begin{equation}
\left\vert \bk{j_z,J_z}{\psi}\right\vert ^{2}=\left\vert \bk{j_x,J_x}{\psi}\right\vert ^{2}=\left\vert \bk{0_y,J_y}{\psi}\right\vert ^{2}=\left\vert \bk{1_y,J_w}{\psi}\right\vert ^{2}=\frac{1}{6} \, .
\label{eq: c6 conditions on psi 2}
\end{equation}

It will take us three steps to show that each of these two sets of equations is contradictory. In other words, there is \emph{no} state $\ket{\psi}$ satisfying either the constraints (\ref{eq: c6 conditions on psi}) or (\ref{eq: c6 conditions on psi 2}).

Given a candidate state $\ket{\psi}\in \mathbb{C}^6$ we will derive (Step 1) that  the smaller subsystem must reside in a totally mixed state 
which implies that the unknown state $\ket{\psi} \in \mathbb{C}^6$ is maximally entangled,
\begin{equation}
\ket{\psi}=\frac{1}{\sqrt{2}}\Bigl(\ket{0_z}\otimes\ket D+\ket{1_z}\otimes\ket{D^{\perp}}\Bigr)\, ,
\label{eq: maxentangled}
\end{equation}
with any two orthogonal states $\ket{D}, \ket{D^\perp} \in\mathbb{C}^{3}$.

Then we will show (Step 2) that the states $\ket{D}$ and $\ket{D^\perp}$ are given by two states either of the basis ${\cal B}_y$ or of ${\cal B}_w$, displayed in Eqs. (\ref{yandwmatrix}). Calling these states $\ket{H}$ and $\ket{H^\perp}$, a total of twelve candidates remains, namely 
\begin{equation}
\ket{\psi}=\frac{1}{\sqrt{2}}
\Bigl( \ket{0_z}\otimes \ket{H} + \ket{1_z}\otimes\ket{H^\perp} \Bigr)\, .  
\label{eq:impossible candidate}
\end{equation}
However, any state $\ket{\psi}$ of the form (\ref{eq:impossible candidate}) will turn out to be incompatible with some MU conditions not used so far (Step 3). 

\textbf{Step 1}: Fix the values of $j$ and $a$ in Eqs. (\ref{eq: c6 conditions on psi}). Summing over $J$ leads to six equations 
\begin{equation}
\sum_{J=0}^{2}\left\vert \bk{j_a,J_a}{\psi}\right\vert
^{2}=\bra{j_a}\Bigl[\mbox{tr}_{B}\kb{\psi}{\psi}\Bigr]\ket{j_a}=\bra{j_a}\hat{\rho}_{A}\ket{j_a}=\frac{1}{2}\,,
\label{eq:trace
over B}\end{equation}
which are sufficient to determine the components of the Bloch vector $\bf n$ of $\hat{\rho}_{A} = ({\hat I}_A + {\bf n} \cdot \hat \sigma)/2$. Since the spin components are given by  $\hat{\sigma}_{a}=\kb{0_a}{0_a}-\kb{1_a}{1_a}$, 
one finds that
\begin{equation}
n_{a}\equiv\mbox{tr}_{A}\bigl[\hat{\sigma}_{a}\hat{\rho}_{A}\bigr]=0\, ,\quad
a=x,y,z \,,
\label{eq: Bloch components}
\end{equation}
which means that the smaller subsystem must reside in the maximally mixed state,
\begin{equation}
\hat{\rho}_{A}\equiv\mbox{tr}_{B}\kb{\psi}{\psi}=\frac{1}{2}\hat{I}_{A}\,.
\label{eq:totally mixed rhoA}
\end{equation}
Summing Eqs. (\ref{eq: c6 conditions on psi 2}) over $J$, with $j$ and $a$ fixed, results in the same six equations $\bra{j_a}\hat{\rho}_{A}\ket{j_a}=1/2$, hence 
Eq. (\ref{eq:totally mixed rhoA}) holds in this case as well. 

Next, the Schmidt decomposition of a state $\ket{\psi}\in\mathbb{C}^{6}$ reads
\begin{equation}
\ket{\psi}=\lambda_{1}\ket c\otimes\ket
C+\lambda_{2}\ket{c^{\perp}}\otimes\ket{C^{\perp}}
\label{eq: general Schmidt decomp}
\end{equation}

where $\bigl\{\ket c,\ket{c^{\perp}}\bigr\}$ and $\bigl\{\ket
C,\ket{C^{\perp}},\ket{C^{\perp\!\!\!\perp}}\bigr\}$
are appropriately chosen orthonormal bases of the spaces $\mathbb{C}^{2}$
and $\mathbb{C}^{3}$ respectively, while $\lambda_{1,2}$ are two
positive numbers satisfying $\lambda_{1}^{2}+\lambda_{2}^{2}=1$. Eq. (\ref{eq:totally mixed rhoA}) implies that these coefficients must be equal so that $\lambda_{1}=\lambda_{2}=1/\sqrt{2}$ follows. Consequently, we are free to identify the basis $\bigl\{\ket c,\ket{c^{\perp}}\bigr\}$ with the standard basis $\bigl\{\ket{0_z},\ket{1_z}\bigr\}$ of $\mathbb{C}^{2}$, at the expense of using a different orthonormal basis $\bigl\{\ket D,\ket{D^{\perp}},\ket{D^{\perp\!\!\!\perp}}\bigr\}$
of $\mathbb{C}^{3}$, unitarily equivalent to $\bigl\{\ket
C,\ket{C^{\perp}},\ket{C^{\perp\!\!\!\perp}}\bigr\}$.
Thus, the candidates for states MU to three product bases
must be maximally entangled ones,
\begin{equation}
\ket{\psi}=\frac{1}{\sqrt{2}}\Bigl(\ket{0_z}\otimes\ket D+\ket{1_z}\otimes\ket{D^{\perp}}\Bigr)\, .
\label{eq: special Schmidt decomp}
\end{equation}
This result agrees with a known result: if a complete set of seven MU bases in dimension six contains three MU product bases then all states of the remaining four MU bases are maximally entangled \cite{wiesniak+11}.
 
\textbf{Step 2}: Now consider the reduced density matrix for the larger subsystem (with label $B$), 
\begin{equation}
\hat{\rho}_{B}= \frac{1}{2} \bigl(\kb DD+\kb{D^{\perp}}{D^{\perp}}\bigr) \, ,
\label{reddmB}
\end{equation}
which has eigenvalues $(1/2,1/2,0)$, in agreement with those of $\hat{\rho}_{A}$ in (\ref{eq:totally mixed rhoA}), except for a padded zero. The requirement that the state $\ket{\psi}$ be MU to the states
$\bigl\{\ket{j_z,J_z}\bigr\}$ and $\bigl\{\ket{j_x,J_x}\bigr\}$, which appear in both triples, imposes restrictions on the states $\ket D$ and $\ket{D^{\perp}}$. Summing the conditions in (\ref{eq: c6 conditions on psi}) and $(\ref{eq: c6 conditions on psi 2})$ over all values of $j$ while keeping $J$ fixed, one obtains six further constraints now on the density matrix $\hat{\rho}_{B}$,
\begin{equation}
\sum_{j=0}^{1}\left\vert \bk{j_a,J_a}{\psi}\right\vert^{2}
=\bra{J_a}\Bigl[\mbox{tr}_{A}\kb{\psi}{\psi}\Bigr]\ket{J_a} 
\equiv \bra{J_a}\hat{\rho}_{B}\ket{J_a}=\frac{1}{3}\,,
\label{eq:trace over A}
\end{equation}
where $J=0,1,2$ and $a=x,z$, similar in spirit to Eqs. (\ref{eq:trace over B}). However, these expectation values are \emph{not} sufficient to reconstruct the reduced density matrix $\hat{\rho}_{B}$. Nevertheless, one can draw the important conclusion that 
\begin{equation}
\bigl|\bk{J_a}{D^{\perp\!\!\!\perp}}\bigr|^{2}=\frac{1}{3}\,, 
\quad J=0,1,2, \quad a = x,z\, .  
\label{eq: MU condition in C3}
\end{equation}
To see this, use the resolution of the identity in terms of the $D$-basis
of $\mathbb{C}^{3}$ to rewrite (\ref{reddmB}) as 
\begin{equation}
\hat{\rho}_{B}=\frac{1}{2}\bigl(\hat{I}_{B}-\kb{D^{\perp\!\!\!\perp}}{D^{\perp\!\!\!\perp}}\bigr) \, 
\label{eq:alternative rho B}
\end{equation}
and calculate its expectation value in the state $\ket{J_a}$.

Eqs. (\ref{eq: MU condition in C3}) tell us that the state
$\ket{D^{\perp\!\!\!\perp}}$ is MU to the states of the MU bases
${\cal B}_x$ and ${\cal B}_z$ of $\mathbb{C}^{3}$. This leaves only a
small number of possibilities for the state
$\ket{D^{\perp\!\!\!\perp}}$: it must coincide with one of the six
vectors $\ket{J_y},\ket{J_w}, J=0,1,2$, which form ${\cal B}_y$ and
${\cal B}_w$, since - as shown in \cite{brierley+10} - these are indeed the 
only states in $\mathbb{C}^{3}$ MU
to the pair ${\cal B}_z$ and ${\cal B}_x$. Letting
$\ket{D^{\perp\!\!\!\perp}} \equiv \ket{H^{\perp\!\!\!\perp}}$, where
$\ket{H^{\perp\!\!\!\perp}}$ is any of the six states in ${\cal B}_y 
\cup {\cal B}_w$, the states $\ket{D}$ and $\ket{D^{\perp}}$ must be
linear combinations of $\ket{H}$ and $\ket{H^\perp}$. After removing 
overall phase factors, we can thus write
\begin{equation}
\begin{array}{rcl}
\ket{D} &=&\cos\frac{\vartheta}{2}\,\ket{H} +e^{i\phi}\sin\frac{\vartheta}{2}\,\ket{H^\perp}\,, \\
\ket{D^{\perp}} &=& \sin\frac{\vartheta}{2}\,\ket{H}-e^{i\phi}\cos\frac{\vartheta}{2}\,\ket{{H}^\perp}\,,
\end{array}
\label{eq:omegapm}
\end{equation}
with two real parameters $\vartheta\in[0,\pi]$, and $\phi\in[0,2\pi)$. 
Projecting the candidate $\ket{D}$ given in (\ref{eq:omegapm})
onto the states $\ket{0_z,J_z}, J=0,1,2$, produces three constraints on the free parameters:
\begin{equation}
\bigl|\bk{J_z}{D}\bigr|^{2}\equiv\bigl|\bra{J_z}\bigl(\cos\frac{\vartheta}{2}\,\ket{H}+e^{i\phi}\sin\frac{\vartheta}{2}\,\ket{{H}^\perp}\bigr)\bigr|^{2}=\frac{1}{3}\:.\label{eq:project on zbasis}
\end{equation}
Using $\bigl|\bk{J_z}{H}\bigr|^{2}=\bigl|\bk{J_z}{{H}^\perp}\bigr|^{2}=1/3$,
this equation leads to the conditions
\begin{equation}
\sin\frac{\vartheta}{2}\,\cos\frac{\vartheta}{2}\,\Bigl(e^{i\phi}\bk{H}{J_z}\bk{J_z}{{H}^\perp}+\mbox{c.c}\Bigr)=\frac{1}{3}\sin\vartheta\,\cos(\phi+\mu_{J})=0 \, ,
\label{eq: trig condition}
\end{equation}
where the relation $\bk{H}{J_z}\bk{J_z}{{H}^\perp} \equiv(1/3)\, e^{i\mu_J}$ defines the angles $\mu_J \in [0, 2\pi), J=0,1,2$. However, the states $\ket{H}$ and $\ket{{H}^\perp}$ are orthogonal, which implies that
\begin{equation}
0=\bk{H}{{H}^\perp}
= \sum_{J=0}^{2}\bk{H}{J_z}\bk{J_z}{{H}^\perp}
= \frac{1}{3}\sum_{J=0}^{2}e^{i\mu_{J}} \, ,
\label{01ortho}
\end{equation}
forcing \begin{equation}
\mu_{J}=\mu+\frac{2\pi}{3}J\:,\quad J=0,1,2\,,
\label{muvalues}
\end{equation}
with some constant $\mu\in[0,2\pi)$. Therefore, Eqs. (\ref{eq: trig condition}) require either $\sin \vartheta \equiv 0$ or
\begin{equation}
\cos(\phi+\mu+\frac{2\pi}{3}J)=0,\: J=0,1,2\, .
\end{equation}
Since the zeros of the cosine function occur at intervals of length $\pi$ (not $2\pi/3$), we conclude that $\vartheta/2\in\left\{0,\pi/2\right\} $ are the only values allowed in (\ref{eq:omegapm}). An entirely analogous argument leads to the same conclusion if we consider the state $\ket{D^\perp}$ defined in (\ref{eq:omegapm}) instead of $\ket{D}$.

Thus, we have shown that there are only two cases in which the requirements of (\ref{eq: c6 conditions on psi}) or (\ref{eq: c6 conditions on psi 2}) are satisfied: we must have either
\begin{equation}
\ket{D} = \ket{H} \quad \mbox{ and }  \quad \ket{D^{\perp}} = - e^{i\phi}\ket{{H}^\perp} \, ,
\label{eq:simpler omegapm1}
\end{equation}
or
\begin{equation}
\ket{D} = e^{i\phi}\ket{{H}^\perp} \quad \mbox{ and }  \quad  \ket{D^{\perp}} = \ket{H} \, .
\label{eq:simpler omegapm2}
\end{equation}
In both cases, the phase factors may be absorbed into the definition of the state $\ket{{H}^\perp}$, which leaves us with two possible candidates being MU to the three product bases in $\mathcal{T}_{0}$ or $\mathcal{T}_{1}$, namely
\begin{equation}
\ket{\psi} = \frac{1}{\sqrt{2}}
\Bigl(\ket{0_z}\otimes\ket{H} + \ket{1_z}\otimes\ket{{H}^\perp}\Bigr) \, , 
\label{eq: two impossible candidates}
\end{equation}
and the state obtained from swapping $\ket{H}$ with $\ket{{H}^\perp}$.
Consequently, the requirement of the state $\ket{D^{\perp\!\!\!\perp}}$  to be a member of ${\cal B}_y$ or ${\cal B}_w$ implies that the states $\ket{D}$ and $\ket{D^\perp}$ must coincide with the two other members of the same basis. Overall, we have indeed reduced the possible states mutually unbiased to $\mathcal{T}_{0}$ or  $\mathcal{T}_{1}$ to twelve entangled states listed in Eq. (\ref{eq:impossible candidate}).

\textbf{Step 3}: Finally, we show that states $\ket{\psi}$ of the form (\ref{eq: two impossible candidates}) are not MU to the states $\ket{1_x,J_x}, J=0,1,2,$ which are present in both product triples, $\mathcal{T}_{0}$ and  $\mathcal{T}_{1}$. The mechanics to produce this contradiction is similar to the one given at the end of Step 2.

To begin, let us consider the state $\ket{\psi}$ in (\ref{eq: two impossible candidates}): the conditions
\begin{equation}
\frac{1}{2}\, \left| \bra{1_x,J_x}
\Bigl(\ket{0_z}\otimes\ket{H} + \ket{1_z}\otimes\ket{{H}^\perp}\Bigr)\rangle \right|^{2}
=\frac{1}{6}\end{equation}
lead to 
\begin{equation}
\bk{H}{J_x}\bk{J_x}{{H}^\perp}+\bk{{H}^\perp}{J_x}\bk{J_x}{H}=0\, .
\end{equation}
Upon writing $\bk{H}{J_x}\bk{J_x}{{H}^\perp}\equiv(1/3)\, e^{i\nu_{J}}$,
one obtains
\begin{equation}
\cos \left(\nu + \frac{2\pi}{3} J \right) = 0 \,,\quad J=0,1,2\,,
\label{eq:second trig equations}
\end{equation}
where we have used the fact that the orthogonality of the states $\ket{H}$ and $\ket{{H}^\perp}$ restricts the values of the phases $\nu_{J}$ in analogy to  Eqs. (\ref{muvalues}). However, the three equations in (\ref{eq:second trig equations}) cannot hold simultaneously, and the state $\ket{\psi}$ in (\ref{eq: two impossible candidates}) is found \emph{not} to be  MU to the given three product bases. The same contradiction occurs for the other eleven states listed in (\ref{eq:impossible candidate}) which completes the proof that there is not a single state mutually unbiased to the triple ${\cal T}_0$ or ${\cal T}_1$. 
 
\section{An unextendible MU product constellation}

A MU \emph{constellation} is a set of states that contains both orthogonal and mutually unbiased states \cite{brierley+08}. MU constellations result, for example, upon removing states from a complete set of MU bases. A constellation which contains only product states will be called a MU \emph{product} constellation.

We now marginally strengthen Theorem \ref{thm:unextendible-triple} by considering the  constellation $\{5,5,4\}^{\otimes}_6$ which consists of two product bases 
(five orthonormal rays in $\mathbb{C}^6$ determine a unique sixth state so that it is not listed in this notation), and a set $\mathcal{S}$ of four orthogonal product states. 
\begin{thm}
\label{theorem:haveno554}
The product constellation $\{5,5,4\}^{\otimes}_6$ cannot be part of a complete set of seven MU bases.
\end{thm}
This result is an immediate consequence of the following lemma, the proof of which will be the main part of this section. 
\begin{lemma} 
\label{lemma:-extending-small-constellation}
The product constellation $\{5,5,4\}^{\otimes}_6$ extends to a triple of MU bases only by adding \emph{product} states.
\end{lemma}
If the product constellation $\{5,5,4\}^{\otimes}_6$ was part of a complete set of seven MU bases, Lemma \ref{lemma:-extending-small-constellation} would imply that the complete set must contain a \emph{triple} of MU product bases, contradicting Theorem \ref{thm:unextendible-triple}.

To prove Lemma \ref{lemma:-extending-small-constellation}, we need the complete list of pairs of MU product bases obtained in \cite{mcnulty+11}:

\begin{lemma}
\label{thm:All-pairs-d=6} Any pair of MU product bases in the
space $\mathbb{C}^{2}\otimes\mathbb{C}^{3}$ is equivalent to a member
of the families \begin{align}
\mathcal{P}_{0} & =\{\ket{j_{z},J_{z}};\,\ket{j_{x},J_{x}}\}\,,
\nonumber \\
\mathcal{P}_{1} &
=\{\ket{j_{z},J_{z}};\,\ket{0_{x},J_{x}},\ket{1_{x},\hat{R}_{\xi,\eta}J_{x}}\}\,, \nonumber \\
\mathcal{P}_{2} &
=\{\ket{0_{z},J_{z}},\ket{1_{z},J_{y}};\,\ket{0_{x},J_{x}},\ket{1_{x},J_{w}}\}\,, \nonumber \\
\mathcal{P}_{3} &
=\{\ket{0_{z},J_{z}},\ket{1_{z},\hat{S}_{\zeta,\chi}J_{z}};\,\ket{j_{x},0_{x}},\ket{\hat{r}_{\sigma}j_{x},1_{x}},\ket{\hat{r}_{\tau}j_{x},2_{x}}\}\,,\end{align}
 with $j=0,1$ and $J=0,1,2$. The unitary operator $\hat{R}_{\xi,\eta}$
is defined as
$\hat{R}_{\xi,\eta}=\kb{0_{z}}{0_{z}}+e^{i\xi}\kb{1_{z}}{1_{z}}+e^{i\eta}\kb{2_{z}}{2_{z}}\,,$
for $\eta,\xi\in[0,2\pi)$, and $\hat{S}_{\zeta,\chi}$ is defined
analogously with respect to the $x$-basis; the unitary operators
$\hat{r}_{\sigma}$ and $\hat{r}_{\tau}$ act on the basis
$\{\ket{j_{x}}\}\equiv\{\ket{\pm}\}$
according to $\hat{r}_{\sigma}\ket{j_{x}}=(\ket{0_{z}}\pm
e^{i\sigma}\ket{1_{z}})/\sqrt{2}$
for $\sigma\in(0,\pi)$, etc.
\end{lemma}

The four product states in $\mathcal{S}$ must be MU to one of the pairs listed in Lemma \ref{thm:All-pairs-d=6}. However, we can exclude the pairs $\mathcal{P}_2$ and $\mathcal{P}_3$ since no product state can be MU to either pair, as follows from a result also derived in \cite{mcnulty+11}: 

\begin{lemma}
\label{lemma:-product-states-MU-to-sets}
The product state $\ket{\phi,\Phi}\in\mathbb{C}^{6}$ is MU to the product basis $\{\ket{\psi_{i},\Psi_{i}}\}$ with $i=1\ldots 6$,
if and only if $\ket{\phi}$ is MU to $\ket{\psi_{i}}\in\mathbb{C}^{2}$
and $\ket{\Phi}$ is MU to $\ket{\Psi_{i}}\in\mathbb{C}^{3}$.
\end{lemma}

The pair $\mathcal{P}_2$ contains a complete set of four MU bases for the space $\C^3$ which means there is no other product state MU to $\mathcal{P}_2$. Similarly, no state in $\C^2$ is MU to the bases $\{\ket{j_z}\}$, $\{\ket{j_x}\}$ and $\{\ket{\hat{r}_\sigma j_x}\}$, ($\hat{r}_\sigma\neq\hat{I})$, and therefore no product state MU to $\mathcal{P}_3$ exists. Thus, the two MU bases of any MU product constellation of the form $\{5,5,4\}^{\otimes}_6$ are given by either of the pairs ${\mathcal P}_0$ or ${\mathcal P}_1$. 

We now use Lemmas \ref{thm:All-pairs-d=6} and \ref{lemma:-product-states-MU-to-sets} to limit the form of the four states which make up the set $\mathcal{S}$. Since there are only three MU bases in $\C^2$, the first factor of each of the four states in $\mathcal{S}$ must be either $\ket{0_y}$ or $\ket{1_y}$, giving rise to only two possibilities, either
\begin{equation}
\mathcal{S}_1 = \{\ket{0_y,A}, \ket{0_y,A^\perp}, \ket{0_y, A^{\perp\!\!\!\perp}}, \ket{1_y, B}\}
\label{constellation1}
\end{equation}
or
\begin{equation}
\mathcal{S}_2 = \{\ket{0_y,A}, \ket{0_y,A^\perp}, \ket{1_y, B}, \ket{1_y, B^\perp}\},
\label{constellation2}
\end{equation}
where $\{\ket{A}, \ket{A^\perp}, \ket{A^{\perp\!\!\!\perp}}\}$ and $\{\ket{B}, \ket{B^\perp}, \ket{B^{\perp\!\!\!\perp}}\}$ denote two orthonormal bases in $\C^3$.
The crucial point here is to observe that both $\ket{0_y}$ and $\ket{1_y}$ can occur at most three times as a factor -- otherwise the states in $\mathcal{S}$ could not be orthogonal. Each state of the set $\mathcal{S}$ must be MU to all states of either $\mathcal{P}_0$ or $\mathcal{P}_1$, which implies that any one of the six states $\ket{A}, \ldots, \ket{B^{\perp\!\!\!\perp}}$, occurring in (\ref{constellation1}) or (\ref{constellation2}) must be MU to the bases ${\cal B}_z$ and ${\cal B}_x$. This requirement limits the states to members of the bases ${\cal B}_y$ or ${\cal B}_w$. 

The states $\ket{1_y, B^\perp}$ and $\ket{1_y, B^{{\perp\!\!\!\perp}}}$ are orthogonal to the quadruple (\ref{constellation1}), as are their linear combinations,
\begin{equation}
\ket{\psi_1} =
\alpha\ket{1_y, B^\perp}+\beta\ket{1_y, B^{{\perp\!\!\!\perp}}}
\equiv \ket{1_y} \otimes (\alpha\ket{B^\perp}+\beta\ket{B^{{\perp\!\!\!\perp}}}) \, , 
\end{equation}
with $|\alpha|^2+|\beta|^2=1$. Hence, adding any two orthogonal states from this family to the set $\mathcal{S}_1$ in (\ref{constellation1}) produces a MU \emph{product} basis. 

Any orthonormal state extending the set $\mathcal{S}_2$ in (\ref{constellation2}) can be written as
\begin{equation}
\ket{\psi_2}=\alpha\ket{0_y,A^{\perp\!\!\!\perp}}+\beta\ket{1_y,B^{\perp\!\!\!\perp}},
\end{equation}
which is \emph{entangled} unless $\ket{A^{\perp\!\!\!\perp}}=\ket{B^{\perp\!\!\!\perp}}$ or one of the constants $\alpha$ and $\beta$ is zero. We now show that the state $\ket{\psi_2}$ cannot be entangled if it is to satisfy the MU conditions
\begin{equation}
|\bk{0_z,J_z}{\psi_2}|^2=\frac{1}{6} \, , \quad \quad J=0,1,2 \, .
\label{MUcondonpsi2}
\end{equation}
Write $\ket{A^{\perp\!\!\!\perp}}=(\omega_0 \ket{0_z}+\omega_1\ket{1_z}+\omega_2\ket{2_z})/\sqrt{3}$ and $\ket{B^{\perp\!\!\!\perp}}=(\omega_0^\prime\ket{0_z}+\omega_1^\prime\ket{1_z}+\omega_2^\prime\ket{2_z})/\sqrt{3}$,
where $\omega_0 = \omega_0^\prime = 1$ and each of the four coefficients $\omega_1, \ldots, \omega_2^\prime$ is a third root of unity such that the states $\ket{A^{\perp\!\!\!\perp}}$  and $ \ket{B^{\perp\!\!\!\perp}}$ coincide with any two different states of the bases $\mathcal{B}_y$ and $\mathcal{B}_w$. Then, the MU conditions in (\ref{MUcondonpsi2}) turn into 
\begin{equation}
\label{eq:alpha-beta}
\alpha \bar\beta \bk{J_z}{A^{\perp\!\!\!\perp}}\bk{B^{\perp\!\!\!\perp}}{J_z}  + \bar\alpha\beta \bk{J_z}{B^{\perp\!\!\!\perp}}\bk{A^{\perp\!\!\!\perp}}{J_z} = 0\, , \quad J = 0,1,2 \, ,
\end{equation}
or explicitly,
\be
\alpha \bar\beta \omega_J{\bar\omega_J}^\prime +\bar \alpha \beta {\bar\omega_J}\omega_J^\prime = 0 \, , \quad J=0,1,2 \, .
\label{explicitMU}
\ee
For $J=0$, we find the relation
\begin{equation}
\alpha \bar\beta + \bar\alpha\beta = 0 \, ,
\end{equation}
which, when used in (\ref{explicitMU}), leads to
\be
\alpha \bar\beta (\omega_J{\bar\omega}_J^\prime-{\bar\omega_J}\omega_J^\prime)=0 \, , \quad J=1,2 \, .
\ee
However, these constraints on the phase factors cannot be satisfied by any allowed choice of the pair of states $\ket{A^{\perp\!\!\!\perp}}$  and $ \ket{B^{\perp\!\!\!\perp}}$ with $\ket{A^{\perp\!\!\!\perp}}\neq\ket{B^{\perp\!\!\!\perp}}$.  Thus, either $\alpha$ or $\beta$ must equal zero, and we conclude that $\ket{\psi_2}$ is a product state. This completes the proof of Lemma \ref{lemma:-extending-small-constellation}.

\section{Concluding Remarks}

The main result of this paper is an \emph{analytical} proof that no vector is MU to any triple of MU product bases, i.e. Theorem \ref{thm:unextendible-triple}. Our approach exploits the structure of MU product bases in a novel fashion, and it is entirely independent of any computer-aided results. Thus, we consider it to be a worthy addition to the few existing analytic results on MU bases in dimension six.

Results stronger than Theorem 1 are known which exclude a wider class of MU bases from complete sets; however, the numerical searches for MU bases with rigorous error bounds \cite{Jaming+2009} and the proof that the Heisenberg-Weyl pair of bases $\{\ket{j_z,J_z}\}$ and $\{\ket{j_x,J_x}\}$ is MU to at most one further basis \cite{grassl+04} rely on  a computer in one way or another. Interestingly, the latter result immediately provides an alternative (computer-aided) proof of Theorem 1: firstly, the pair $\{\ket{j_z,J_z}\}$ and $\{\ket{j_x,J_x}\}$ is present in both product triples $\mathcal{T}_0$ and $\mathcal{T}_1$ and, secondly, any product triple is known to be equivalent to one of these two triples.

A recent analytic result \cite{matolcsi+12} employs combinatorial and Fourier analytic arguments to prove that no complete set of MU bases in dimension six will contain both the standard and Fourier basis. As a consequence, no complete set of MU bases will contain triples of MU product bases. Whilst this is also a consequence of Theorem \ref{thm:unextendible-triple}, the result presented here is different to the result in \cite{matolcsi+12} since we have shown the impossibility to extend a product triple by a \emph{single} MU vector. The result in \cite{matolcsi+12} does not seem to forbid such an extension.
  
In order to strengthen Theorem \ref{thm:unextendible-triple} we also considered a MU product \emph{constellation} that is slightly smaller than MU product triples. The resulting Theorem 2 states that the product constellation $\{5,5,4\}^\otimes_6$ cannot be part of a complete set of seven MU bases. Its derivation relies on an enumeration of \emph{all} pairs of MU product bases in dimension six which was given in \cite{mcnulty+11}. 

To make any stronger statements regarding product constellations seems to be surprisingly difficult. For example, we are not able to show whether a complete set of seven MU bases may (or may not) contain the MU constellation $\{5,4,4\}^\otimes_6$, consisting of one MU product basis and two sets of four orthogonal MU product states. The main difficulty is that the proof of Theorem \ref{theorem:haveno554} relies on Lemma \ref{lemma:-product-states-MU-to-sets} which does not apply to the case of the constellation $\{5,4,4\}^\otimes_6$. 

It is worth recalling that a hypothetical complete set of MU bases in dimension six will contain at most \emph{one} product basis \cite{weigert+11}. While this is a stronger statement than the one obtained here, the proof depends on a numerical search with rigorous error bounds \cite{Jaming+2009}. We hope that the analytic proof given here will be extended to cover this stronger result and, ultimately, will help put to rest the existence question of complete sets of MU bases in composite dimensions -- or at least in dimension six, for a start. 

\subsubsection*{Acknowledgements}
We thank Maurice Kibler for carefully reading a draft of this paper. This work has been supported by EPSRC.

\end{document}